\input{aipcheck}

\documentclass[
    ,final            
  ]
  {aipproc}

\usepackage{amsmath}

\newcommand{\al}[0]{\ensuremath{\alpha}}
\newcommand{\ep}[0]{\ensuremath{\varepsilon}}

\newcommand{\si}[0]{\ensuremath{\sigma}}
\newcommand{\be}[0]{\ensuremath{\beta}}

\layoutstyle{6x9}

\begin{document}

\title{Fluctuations, Saturation, and Diffractive Excitation in High Energy 
Collisions\footnote{Work supported in part by the Marie
    Curie RTN ``MCnet'' (contract number MRTN-CT-2006-035606).}}


\author{Christoffer Flensburg\footnote{\ \ In collaboration with G\"osta Gustafson and Leif L\"onnblad}}{
  address={ Dept.~of Theoretical Physics,
    S\"olvegatan 14A, S-223 62  Lund, Sweden}
}

\begin{abstract}
Diffractive excitation is usually described by the Good--Walker 
formalism for low masses, and by the triple-Regge
formalism for high masses. In the Good--Walker formalism the cross section is
determined by the fluctuations in the interaction. By taking the fluctuations in the BFKL ladder into account, it is possible to
describe both low and high mass excitation in the Good--Walker formalism.
In high energy $pp$ collisions the fluctuations are strongly suppressed by 
saturation, which implies that pomeron exchange does not factorise between DIS
and $pp$ collisions. The Dipole Cascade Model reproduces the expected
triple-Regge form for the bare pomeron, and the triple-pomeron coupling is
estimated.
\end{abstract}



\maketitle


\subsection{Introduction}

Diffractive excitation represents large fractions of the cross sections in
$pp$ collisions or DIS. In most analyses of $pp$ collisions low mass excitation is described by
the Good--Walker formalism \cite{Good:1960ba}, while high mass excitation is
described by a triple-Regge formula \cite{Mueller:1970fa, Detar:1971gn}. 
In the Good--Walker formalism the fluctuations in the pomeron ladder 
are normally not included, which is what limits the application to low masses. In the dipole cascade model \cite{Avsar:2005iz,Avsar:2006jy,Avsar:2007xg,Flensburg:2008ag,Flensburg:2010kq} the fluctuations in the ladders are taken into account, allowing the Good--Walker formalism to describe diffractive excitation at all masses.

It turns out that saturation plays a very important role in suppressing diffractive excitation. This means that $pp$ collisions will have a much lower fraction of diffractive excitation than without saturation, specially at higher energies. For DIS, saturation is a smaller effect, and diffractive excitation can be expected to be stronger, which is confirmed by experiments. The impact parameter profile for diffractive excitation in high energy $pp$ is found to be in the shape of a ring, due to the approaching black disc limit at low $b$.

Triple-Regge without saturation predicts powerlike growth with energy of the total, elastic and diffractive excitation cross sections. By removing saturation effects from the dipole cascade model, also that gives a powerlike energy growth, just like the triple-Regge models. The intercept, slope and triple-Pomeron couplings can be extracted from the energy dependencies.

\subsection{The dipole cascade model}
In the Good--Walker formalism the incoming mass eigenstates are not necessarily eigenstates of diffractive interaction. However, the mass eigenstates $\Psi_k$ are linear combinations of the diffraction eigenstates $\Phi_n$ with eigenvalues $T_n$ and coefficients $c_{kn}$. The diffractive cross section of the incoming mass eigenstate $\Psi_0$ can then be written by summing over outgoing mass eigenstates:

$$d\sigma_{\mathrm{diff}}/d^2b = \sum_k \left( \langle \Psi_0 | T | \Psi_k \rangle \right)^2 = \langle \Psi_0 | T^2 | \Psi_0 \rangle = \langle T^2 \rangle$$
where the last average over $T$ is with the weights from $\Psi_0$. By subtracting the elastic part, one finds the diffractive excitation:

$$d\sigma_{\mathrm{diff ex}}/d^2 b  = d\sigma_{\mathrm{diff}}/d^2 b - d\sigma_{\mathrm{el}}/d^2 b =
\langle T^2 \rangle - \langle T \rangle ^2\equiv V_T.$$

It turns out that it is the fluctuations in the interaction amplitude that gives the diffractive excitations.

In our model, the eigenstates of interaction are cascades of colour dipoles in transverse space. Mueller showed that the cascade was equivalent with leading logarithm BFKL \cite{Mueller:1993rr, Mueller:1994jq, Mueller:1994gb} and has since been enhanced with several non-leading order effects such as energy-momentum conservation, confinement, running $\alpha_s$ and improved saturation. The saturation in the interaction is included through unitarisation $T = 1 - e^{-F}$, where $F$ is the Born amplitude. In the cascade, a saturating $2 \rightarrow 2$ ``dipole swing'' is included, providing a saturation in the cascade equivalent to the interaction up to a few percent.

This model has proven to describe a wide range of total, elastic and diffractive cross sections in both $pp$ collisions and DIS. This is described in detail in \cite{Avsar:2007xg,Flensburg:2008ag,Flensburg:2010kq}.

\subsection{Fluctuations and saturation}
\begin{figure}
\includegraphics[angle=270, scale=0.5]{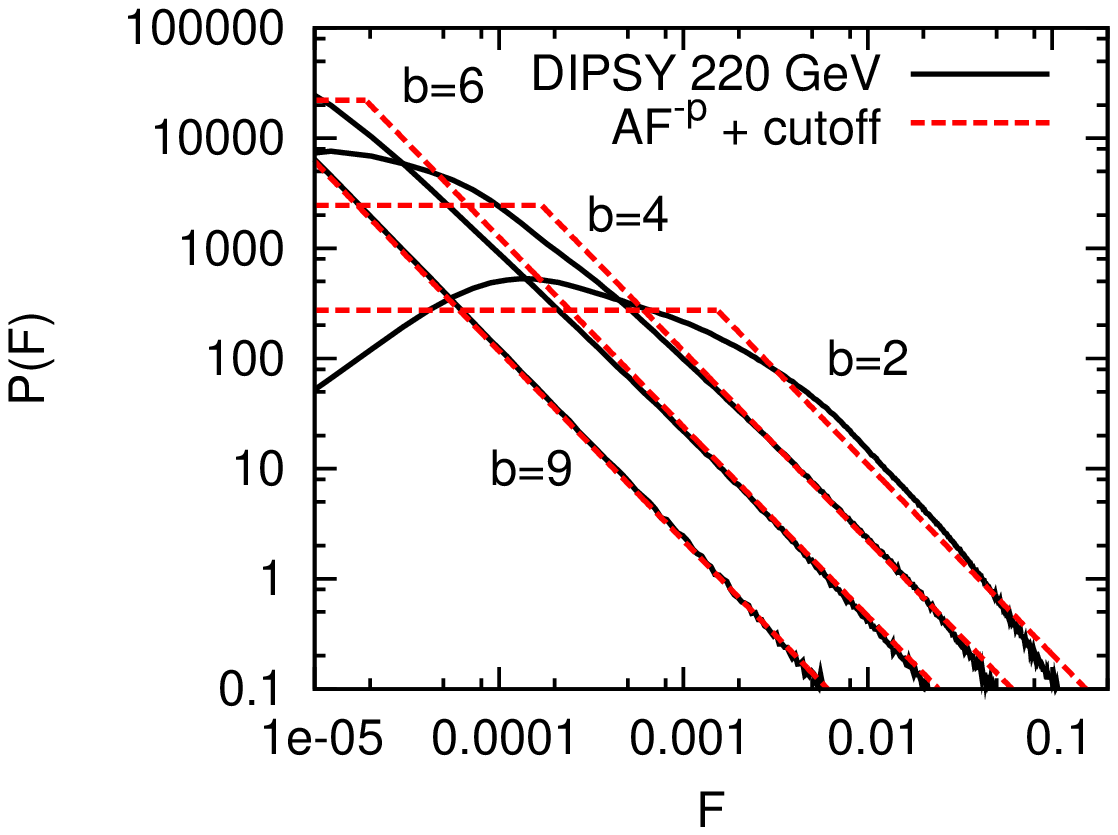}
\hspace{-0.8cm}
\includegraphics[angle=270, scale=0.5]{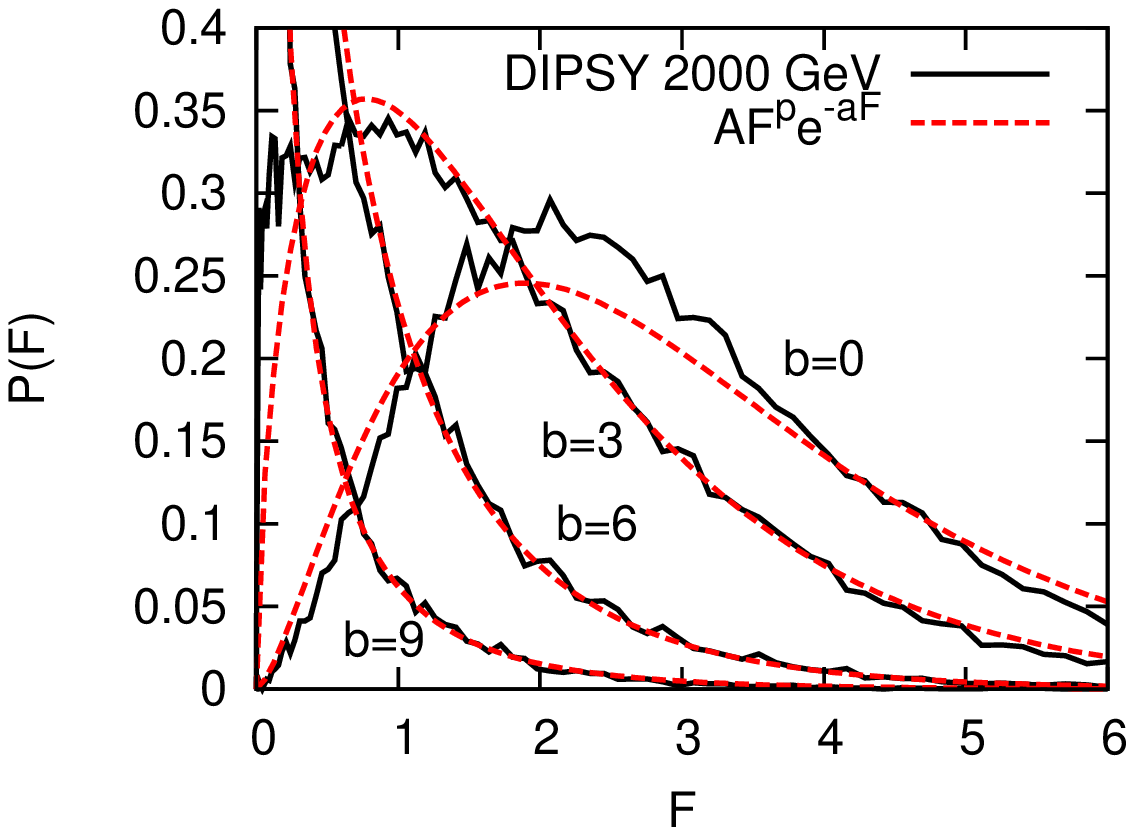}
\hspace{-0.8cm}
\includegraphics[angle=270, scale=0.5]{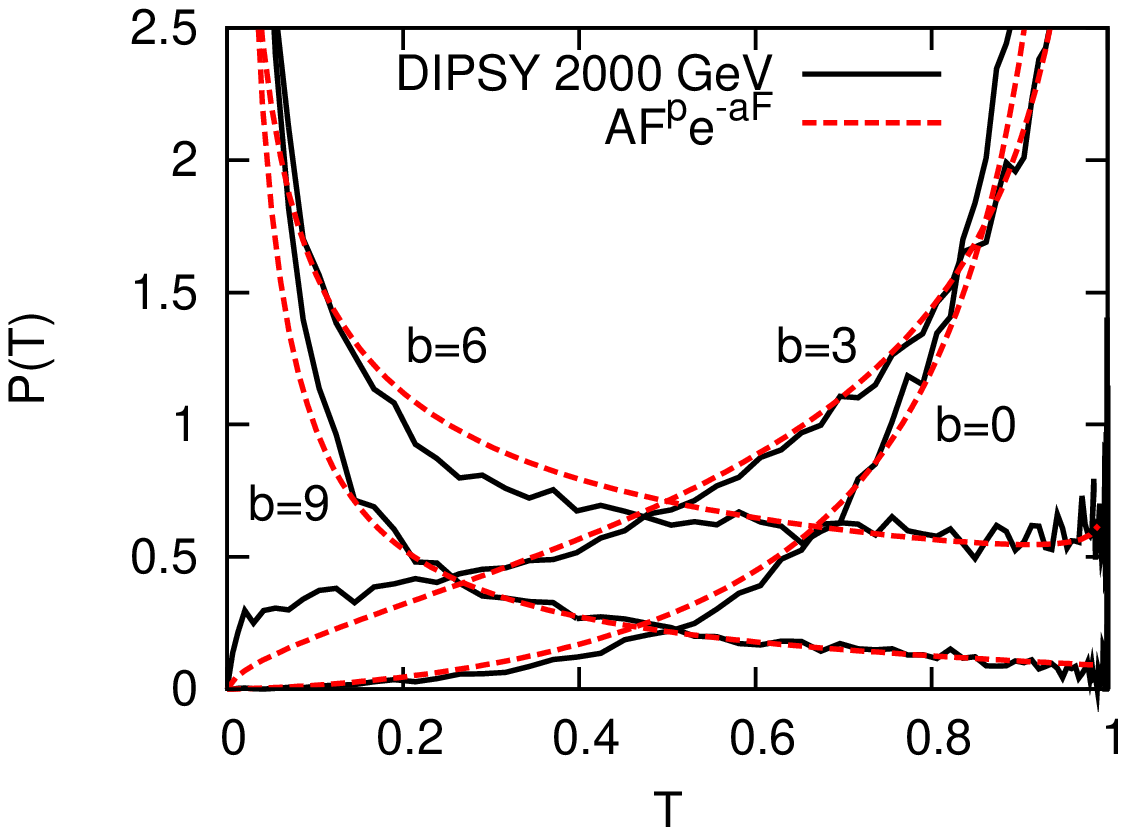}
\caption{\label{fig:flucts} The distribution of interaction amplitudes for DIS (left), unsaturated $pp$ (mid) and saturated $pp$ (right).}
\end{figure}
In this section we will use our model to study how saturation affects the fluctuations in the interaction amplitude, and thus diffractive excitation. We want to study the effect of saturation, and will separate between the Born-level amplitude $F$, and the fully saturated amplitude $T$. Both these are calculated from our model with a Monte Carlo simulation program called DIPSY. A large number of colliding dipole cascades are generated and collided at fix impact parameter, and the frequencies of interaction amplitudes, $P(F)$ and $P(T)$, are studied.

In the left plot of figure \ref{fig:flucts} is the distribution of Born amplitudes for $\gamma^\star p$ at $W = 220$~GeV. It is seen that the distribution behaves roughly as a power of $F$, giving a rather wide distribution and large fluctuations. This corresponds to a large cross section for diffractive excitation in DIS. Since $F$ is well below unity, $T \approx F$ and saturation is a small effect.

\begin{figure}
\includegraphics[angle=270, scale=0.45]{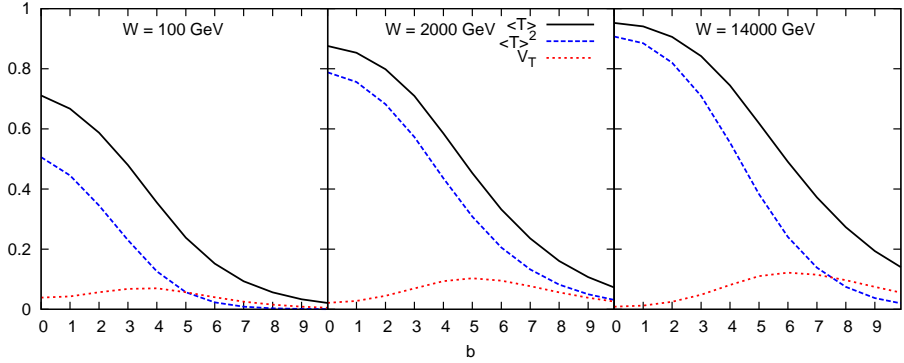}
   \caption{\label{fig:IPP}Impact parameter distributions from the MC for 
    $\langle T\rangle=(d\sigma_{\mathrm{tot}}/d^2b)/2$, $\langle T\rangle^2=d\sigma_{\mathrm{el}}/d^2b$, and 
    $V_T=d\sigma_{\mathrm{diff ex}}/d^2b$ in
    $pp$ collisions at $W=100$, 2000, and 14000 GeV. $b$ is in units of GeV$^{-1}$. }
\end{figure}

The distribution of the Born amplitudes of $pp$ at $\sqrt{s} = 2000$~GeV is shown in the middle plot of figure \ref{fig:flucts}. This distribution behaves as a gamma function, which very wide and corresponds to a very large cross section for diffractive excitation. However, since $F$ is not smaller than 1, unitarity is important. The distribution in the saturated amplitude $T$ is shown in the rightmost plot. The shape for the large $b$ distributions does not change much, but the low $b$ distributions that previously were very wide are now sharply peaked just below $T = 1$. $\langle T \rangle$ approaching 1 corresponds to the black disc limit, and is seen to strongly suppress the fluctuations, and thus the diffractive excitations. This is clearly seen in the impact parameter profile in figure \ref{fig:IPP} where the central collisions get suppressed fluctuations as energy increase. Thus, the diffractive excitations live in a ring where $T \approx 0.5$.

\subsection{Comparison between Good--Walker and triple-Regge}
In unsaturated Regge formalism, the cross sections are
\begin{eqnarray}
\si_{\mathrm{tot}}&=&\be^2(0)s^{\al(0)-1} \equiv \si_0^{p\bar{p}}s^\ep,\nonumber\\
\frac{d\si_{\mathrm{el}}}{dt} &=& \frac{1}{16\pi}\be^4(t)s^{2(\al(t)-1)}, \nonumber\\
M_{\mathrm{X}}^2 \frac{d\si_{\mathrm{SD}}}{dtd(M_{\mathrm{X}}^2)} &=& \frac{1}{16\pi}\be^2(t)\be(0)g_{3\mathrm{P}}(t)
\left( \frac{s}{M_{\mathrm{X}}^2} \right)^{2(\al(t)-1)}\left( M_{\mathrm{X}}^2 \right)^{\epsilon}.
\label{eq:barepomeron}
\end{eqnarray}
Here $\al(t)=1+\ep+\al ' t$ is the 
pomeron trajectory, and $\beta(t)$ and $g_{3P}(t)$ are the proton-pomeron and
triple-pomeron couplings respectively. These cross sections are in most models increasing much faster than the measured ones, and saturation in different forms are added to fit with experiments. To make a comparison between our model in the Good--Walker formalism and the Regge models, it is better to compare the unsaturated models, to avoid the model dependence in the saturation scheme.

\begin{figure}
\includegraphics[width=0.35\linewidth,angle=270]{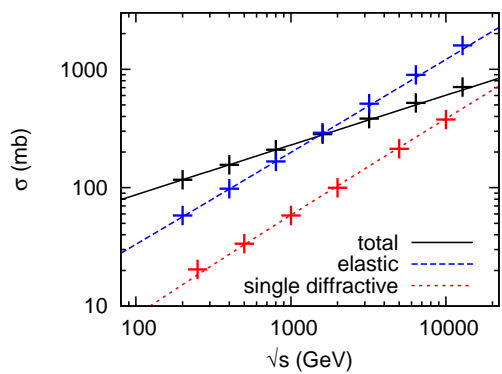}
  \caption{\label{fig:GWvsTR}The total, elastic and single diffractive cross
    sections in the \emph{one-pomeron} approximation. The crosses are from the 
    dipole cascade model without saturation,
    and the lines are from a tuned triple Regge parametrisation.}
\end{figure}

Running the DIPSY Monte Carlo without saturation, it turns out (figure \ref{fig:GWvsTR}) that the energy dependence of the total, elastic and diffractive cross sections fit perfectly to the Regge parametrisations with
\begin{eqnarray}
\al(0)&=&1+\epsilon =1.21, \,\,\,\,\al' = 0.2\,\mathrm{GeV}^{-2},\nonumber\\
\si_0^{p\bar{p}}&=&\beta^2(0)=12.6\,\mathrm{mb}, \,\,\,\, 
b_{0,\mathrm{el}} = 8\,\mathrm{GeV}^{-2}, \,\,\,\, g_{3\mathrm{P}}(t) =\mathrm{const.} = 0.3\,\mathrm{GeV}^{-1}.
\label{eq:parameters}
\end{eqnarray}
This is not a trivial result. For example without the logarithmic corrections in \ref{eq:barepomeron} the fit would have been significantly worse. Similarly, without the confinement, energy conservation and running $\al_s$, that is just leading logarithm BFKL, the increase with energy would have been too strong. The NLL corrections are necessary for the two approaches to agree this well.



\bibliographystyle{aipproc}   

\bibliography{sample,/home/william/people/leif/personal/lib/tex/bib/references,../../Dipole/refs}

\IfFileExists{\jobname.bbl}{}
 {\typeout{}
  \typeout{******************************************}
  \typeout{** Please run "bibtex \jobname" to optain}
  \typeout{** the bibliography and then re-run LaTeX}
  \typeout{** twice to fix the references!}
  \typeout{******************************************}
  \typeout{}
 }

\end{document}